\documentclass{aa}  

\usepackage{graphicx}
\usepackage{txfonts}
\usepackage{lipsum}
\usepackage{xspace}
\usepackage{todonotes}
\usepackage{subcaption}         
\usepackage{lscape}             
\usepackage{placeins}           
 \usepackage{orcidlink}                               

\newcommand{{\hess}}{{H.E.S.S.\xspace}}
\newcommand{{\xrt}}{XRT\xspace}
\newcommand{{\isgri}}{ISGRI\xspace}
\newcommand{{\psrb}}{{ PSR~B1259-63/LS~2883 \xspace}}

\begin{document}

   \title{\textit{INTEGRAL/ISGRI} post 2024-periastron view of PSR B1259-63 }
   \titlerunning{2025 \isgri view of \psrb}
   \authorrunning{Kuzin et al.}


   \author{A. Kuzin\inst{\ref{in:UT} }\fnmsep\thanks{aleksei.kuzin@astro.uni-tuebingen.de}
        \and D. Malyshev\inst{ \ref{in:UT}   \orcidlink{0000-0001-9689-2194}}
        \and M.~Chernyakova\inst{\ref{in:DCU}\orcidlink{0000-0002-9735-3608} }
        \and B.~van~Soelen\inst{\ref{in:UFS}}
        \and A. Santangelo~\inst{\ref{in:UT} }
        }

   \institute{Institut für Astronomie und Astrophysik Tübingen, Universität Tübingen, Sand 1, D-72076 Tübingen, Germany \label{in:UT}
   \and
School of Physical Sciences and Centre for Astrophysics \& Relativity, Dublin City University, Glasnevin, D09 W6Y4, Ireland \label{in:DCU} 
\and
Department of Physics, University of the Free State, PO Box 339, Bloemfontein 9300, South Africa \label{in:UFS}
   }

   \date{Received }

 
   \abstract
   { PSR B1259-63/LS 2883 is a well-studied gamma-ray binary hosting a pulsar in a 3.4-year eccentric orbit around a Be-type star. Its non-thermal emission spans from radio to TeV energies, exhibiting a significant increase near the periastron passage. This paper is dedicated to the analysis of INTEGRAL observations of the system following its last periastron passage in June 2024.
   }
    {We aim to study the spectral evolution of this gamma-ray binary in the soft (0.3-10 keV) and hard (30-300~keV) X-ray energy bands. 
    } 
   {We performed a joint analysis of the data taken by \textit{INTEGRAL}/\isgri in July-August 2024 and quasi-simultaneous \textit{Swift}/\xrt observations. }
   {The spectrum of the system in the 0.3-300~keV band is well described by an absorbed power law  with a photon index of $\Gamma=1.42\pm0.03$. 
   We place constraints on potential spectral curvature,
   limiting the break energy $E_\mathrm{b}>30$~keV for $\Delta\Gamma>0.3$ and cutoff energy $E_\mathrm{cutoff}>150$~keV at 95\% confidence level. 
   For one-zone leptonic emission models, these values correspond to electron distribution spectral parameters of $E_\mathrm{b,e}>0.8$ TeV and $E_\mathrm{cutoff,e}>1.7$ TeV, consistent with previous constraints derived by \hess}   
   {}


   \maketitle

\section{Introduction}
PSR B1259-63 is a rotation-powered radio pulsar with a spin period of $P_{\mathrm{spin}} \approx 48$ ms, orbiting an O9.5V optical companion, LS~2883. 
This binary belongs to the subclass of gamma-ray binaries -- high mass binary systems which 
emit most of their non-thermal emission in the gamma-ray range. 
The system has well-established orbital parameters \citep{John92, MiJ18}, including a high eccentricity ($e \approx 0.87$) and a long orbital period ($T_{\mathrm{orb} }\approx 1237$ days). It is one of only three gamma-ray binaries in which the compact object is confirmed to be a pulsar, making it an attractive candidate for studying and modeling emission from gamma-ray binaries \citep[see][for a review]{ChM20} 

The optical Be star companion produces 
both polar and axisymmetric outflows: a polar wind and a decretion disk \citep{John92}. The pulsar orbital plane is inclined relative to the decretion disk, so the pulsar passes through the disk twice per orbit, approximately $15-20$ days before and after periastron \citep{MJM95}. 

The interaction between the stellar outflows and the pulsar’s relativistic wind forms an intrabinary shock, where accelerated particles produce non-thermal emission spanning from radio to TeV energies \citep{TAK94, Che20}. Close to the periastron passage, this unpulsed emission from the intrabinary shock dominates across the electromagnetic spectrum (from X-rays to TeVs, and in the radio band). For the purpose of our work, we will refer to all such unpulsed, binary-system-generated emission as "binary radiation" (distinguishing it from direct, pulsed emission from a pulsar). The non-thermal radiation intensity increases near periastron, and the X-ray emission is additionally enhanced during the pulsar-disk passages, leading to a two-peaked X-ray light curve. The most recent periastron passage occurred on June 30, 2024 ($T_\mathrm{p} = $~MJD 60491.591).

The multiwavelength observational campaigns covering previous periastron passages have revealed that the system's observational properties, such as the shape of the light curves in different energy ranges, vary from one orbit to another \citep[see][for a review of previous periasters observations]{Che21}. 
In particular, the 2021 periastron passage has revealed a number of unusual features, like the presence of the third X-ray peak \citep{Che21,Che24}. The 2024 periastron was monitored by radio, optical, and gamma-ray observatories, as well as by the X-ray telescopes \textit{Swift} and \textit{NICER}. The X-ray light curve was found to be more complex than expected, demonstrating a prolonged second hump after the second pulsar-disc passage \citep{Che25}. 

In previous studies of the hard X-ray emission from \psrb, the spectrum was successfully described by an absorbed power law (PL) model with no indications of a more complicated spectrum or an additional thermal component. \citet{Shaw04} analyzed the system in the $20-200$ keV band with \textit{INTEGRAL} and found that the PL with an index of $\Gamma = 1.3 \pm 0.3$ provided a good fit. More recently, \citet{Che15} examined data from \textit{NuSTAR} and \textit{Suzaku} (up to $E \sim 70$ keV) as well as \textit{INTEGRAL} (up to $60$ keV, though no spectrum was provided), finding a slightly phase-dependent PL fit with $\Gamma \approx 1.5-1.7$.

 In this work, we examine the hard X-ray emission ($>30$ keV) from the binary using new observations from the \textit{INTEGRAL}/ISGRI observatory. Nominally, the IBIS/ISGRI instrument can obtain X-ray spectra up to $500$ keV, allowing us to test the PL model at higher energies. We analyze data from July-August 2024 to extract the spectrum, build a light curve, and study the spectral evolution of the system. 

The data analysis process is described in Section \ref{sec:data}, followed by results and discussion in Section \ref{sec:results}.  The findings are summarized in Section \ref{sec:concl}.

\section{Data analysis}\label{sec:data}

\begin{table*}[h!]
\caption{\label{t8}Best-fit parameters for the 9 datasets of \xrt--\isgri data, including observation details (ScW ranges, MJD, and time since periastron passage), along with the absorbed (PL) fit parameters $\Gamma$ and $N_\mathrm{H}$ and fluxes in the soft and hard X-ray energy ranges.}
\centering
\begin{tabular}{c c c c c c c c}      
\hline\hline               
Dataset no. & ScW range & MJD &  $T - T_\textrm{p}$ & $F \pm \Delta F \,(0.3-10 \,\mathrm{keV})$ & $F \pm \Delta F\, (30-50 \,\mathrm{keV})$ & $\Gamma \pm \Delta \Gamma$ & $N_\mathrm{H} \pm \Delta N_\mathrm{H}$ \\   
& &  & days &
$10^{-11}~\mathrm{erg}~\mathrm{s}^{-1}~\mathrm{cm}^{-2}$&
$10^{-11}~\mathrm{erg}~\mathrm{s}^{-1}~\mathrm{cm}^{-2}$&
&
$10^{22} \mathrm{cm}^{-2}$\\
\hline
1 & 27980036-60 & 60501.2 & 9.6 &
    $0.87 \pm 0.14$ &
    $0.36 \pm 0.3$ &
    $1.85 \pm 0.72$ & $1.39 \pm 0.87$ \\
    
2 & 27990019-41 & 60503.1 & 11.5 &
    $1.65 \pm 0.19$ &
    $0.93 \pm 0.45$ &
    $1.64 \pm 0.66$ & $0.55 \pm 0.45$ \\
    
3 & 28000051-75 & 60506.5 & 14.9 &
    $3.0 \pm 0.19$ &
    $1.73 \pm 0.4$ &
    $1.66 \pm 0.22$ & $0.87 \pm 0.26$ \\
    
4 & 28010036-60 & 60509.0 & 17.4 &
    $6.08 \pm 0.44$ &
    $4.32 \pm 0.42$ &
    $1.57 \pm 0.11$ & $0.88 \pm 0.37$ \\
    
5 & 28020054-78 & 60511.7 & 20.1 &
    $3.96 \pm 0.15$ &
    $3.93 \pm 0.37$ &
    $1.41 \pm 0.09$ & $0.78 \pm 0.19$ \\
    
6 & 28030030-54 & 60513.8 & 22.2 &
    $4.18 \pm 0.21$ &
    $4.45 \pm 0.53$ &
    $1.38 \pm 0.1$ & $0.82 \pm 0.25$ \\
    
7 & 28040022-48 & 60516.5 & 24.9 &
    $3.11 \pm 0.14$ &
    $3.42 \pm 0.44$ &
    $1.36 \pm 0.1$ & $0.69 \pm 0.18$ \\
    
8 & 28050056-81\tablefootmark{a} & 60519.8 & 28.2 &
    $3.88 \pm 0.18$ &
    $4.6 \pm 0.4$ &
    $1.35 \pm 0.08$ & $1.03 \pm 0.28$ \\
    
9 & 28070023-47 & 60524.5 & 32.9 &
    $5.78 \pm 0.43$ &
    $5.23 \pm 0.49$ &
    $1.48 \pm 0.1$ & $1.25 \pm 0.52$ \\
\hline                                  
\end{tabular}
\tablefoot{All given errors are half sums of lower and upper $1\sigma$ uncertainties.\\
\tablefoottext{a}{Except ScW \#28050079}
}
\end{table*}

The \textit{INTEGRAL}/ISGRI 2024 periastorn passage data covers revolutions 2798-2805 and 2807, with a total exposure time of 258~ks taken between July 9 -- Aug 2 2024. For each revolution, the data were processed according to the IBIS Analysis User Manual v. 11.2\footnote{\url{https://www.isdc.unige.ch/integral/analysis}}, using the standard ISDC offline scientific analysis software v. 11.2. 
Using the standard data analysis pipeline, we obtained nine \isgri spectra for the nine revolutions, summing science windows for each revolution. The energy bins were defined as four logarithmically spanned bins in the $30-80$ keV range and one bin in the $80-300$ keV range.

We note that the spectral analysis of PSR B1259-63/LS~2883 in hard X-rays with \isgri is complicated by the presence of a nearby ($\sim 10'$ away) source 2RXP~J130159.6-635806, a Be/X-ray pulsar, which lies well within the \isgri PSR for our target 
To account for the possible contribution of this source to the PSR B1259-63 spectrum, we utilised the best-fit model from the most recent \textit{NuSTAR} observations of this source taken quasi-simultaneously with \isgri and reported by \citet{Sal25}. However, in 2024, the flux of 2RXP~J130159.6-635806 remained an order of magnitude lower than that of \psrb and does not strongly impact the results.

\begin{figure}[h]
   \centering
   \includegraphics[width=\hsize]{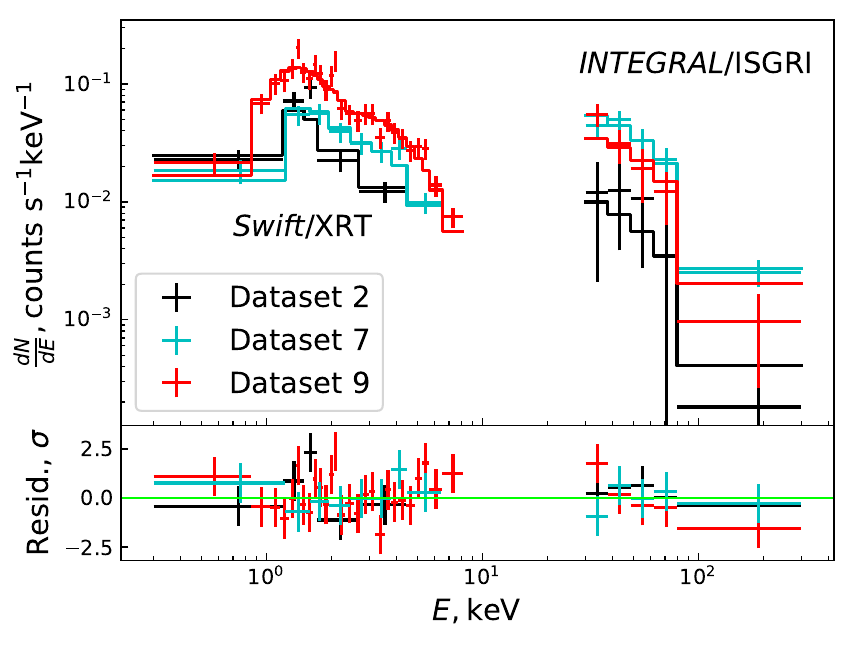}
      \caption{Spectra for the datasets 2, 7, and 9. For each dataset, \textit{Swift}/XRT and \textit{INTEGRAL}/ISGRI data were modeled simultaneously with an absorbed PL model \texttt{tbabs*powerlaw}, and the model is also shown.  }
         \label{fig:spec_3}
   \end{figure}

To extend our analysis to softer X-ray energies, we also incorporated \textit{Swift}/\xrt \citep[hereafter XRT,][]{XRT05} data in the 0.3-10~keV band, as analysed in~\citet{Che25}. The excellent time coverage of \psrb by \xrt allowed us to match each of the nine \isgri observations with a corresponding \xrt observation, separated by no more than 1 day. This resulted in nine joint \xrt and \isgri datasets. The \xrt spectra were binned using a minimum of 25 photons with the \texttt{grppha} routine. These spectra were further analyzed using \texttt{XSPEC v. 12.11.0} from \texttt{HEASOFT v. 6.27}.  Table \ref{t8} shows the parameters of the datasets: Science Window (ScW) ranges, MJD, and time since periastron passage, along with the absorbed PL best fit parameters and fluxes in soft and hard X-ray energy bands.  Figure \ref{fig:spec_3} presents examples of such datasets (no. 2, 7, and 9  ), together with their best fit by the absorbed PL models \texttt{Tbabs*powerlaw} and residuals.

We also searched for indications of a more complex spectral model: broken PL or exponential cutoff PL (see below). For these models, we used the \texttt{XSPEC} models \texttt{Tbabs*bknpower} and \texttt{Tbabs*cutoffpl}. When parameter exploration was required, we employed the \texttt{steppar} routine.

\section{Results and discussion}\label{sec:results}
   \begin{figure}[h]
   \centering
   \includegraphics[width=\hsize]{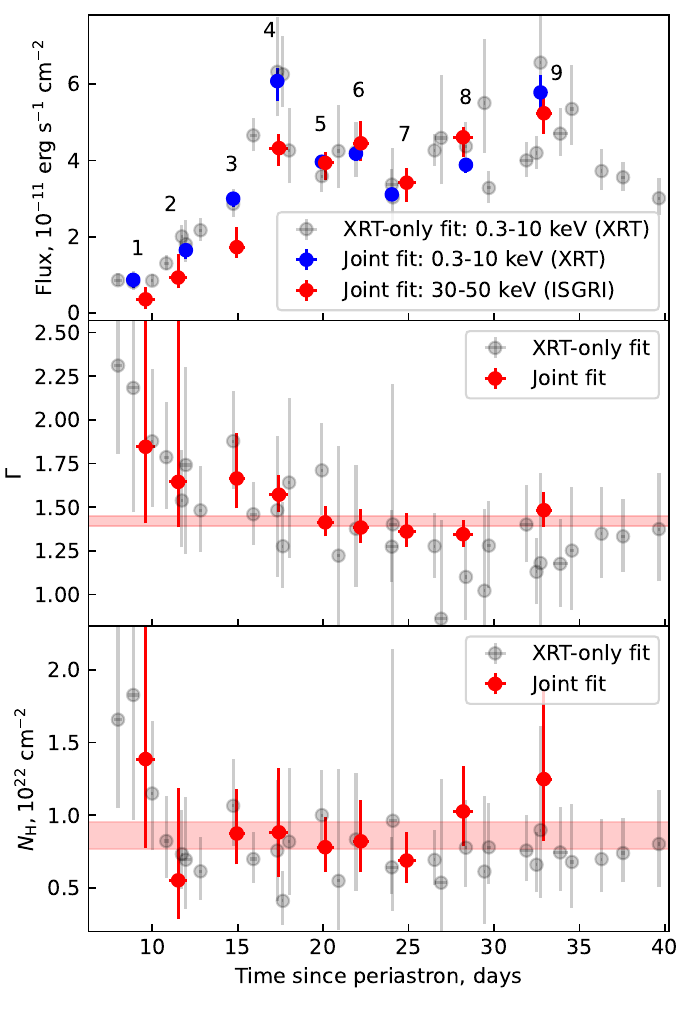}
      \caption{Top: the 2024 light curve in soft X-rays (blue) and hard X-rays (red) from joint \textit{Swift}/XRT and \textit{INTEGRAL}/ISGRI fits. The gray points correspond to the soft X-ray flux from the \xrt-only fit. All \xrt+~\isgri pairs are marked with the dataset number they belong to.  Middle panel (bottom panel): the photon index $\Gamma$ (the column density $N_\mathrm{H}$) from the joint fit, in red, and \xrt-only fit, in gray. On the two lower panels the red bands show the $1\sigma$-uncertainties for the $\Gamma$ and $N_\mathrm{H}$ fits from datasets 4--9. }
         \label{fig:LC_noTh}
   \end{figure}

We fitted the spectra from nine joint (\xrt + \isgri) datasets with the absorbed PL model, treating the interstellar hydrogen density $N_\mathrm{H}$, the X-ray spectral slope $\Gamma$, and the normalisation of the PL as free parameters. For the fit, we adopted $\chi^2$ statistics, the errorbars shown at Fig. \ref{fig:spec_3} correspond to the $1\sigma$ statistical confidence interval.

Fig.~\ref{fig:LC_noTh} presents the results of the performed analysis. 

The top panel illustrates the post-periastron \psrb  light curve obtained from, first only the \xrt data (0.3-10~keV, grey points), and, second, from the joint fit of the \xrt and \isgri data (blue and red points).
The blue points indicate the flux derived from the joint fit in the \xrt energy range (0.3-10~keV), allowing a direct comparison with \xrt-only data points. The red points correspond to the flux in the 30-50~keV energy range.  

The middle panel shows the results for the spectral slope $\Gamma$ derived from \xrt-only data (grey points) and from the joint \xrt-\isgri spectral fit (red points). The lower panel presents the interstellar neutral hydrogen absorption value with the same color scheme. Although not statistically significant, we  note the gradual hardening of the spectral slope as \psrb moves away from periastron. The observed trend is consistent with high-quality XMM-Newton data \citep[see][for the summary for several periastron passages]{Che15}.

Figure~\ref{fig:LC_noTh} illustrates the substantial improvement in the spectral PL  index constrains using the joint \xrt and \isgri data analysis. To further illustrate this improvement, we examined the confidence ellipses for $N_\mathrm{H}$ and $\Gamma$, as these parameters are typically strongly correlated for the X-ray data.

\begin{figure}[h]
   \centering
   \includegraphics[width=\hsize]{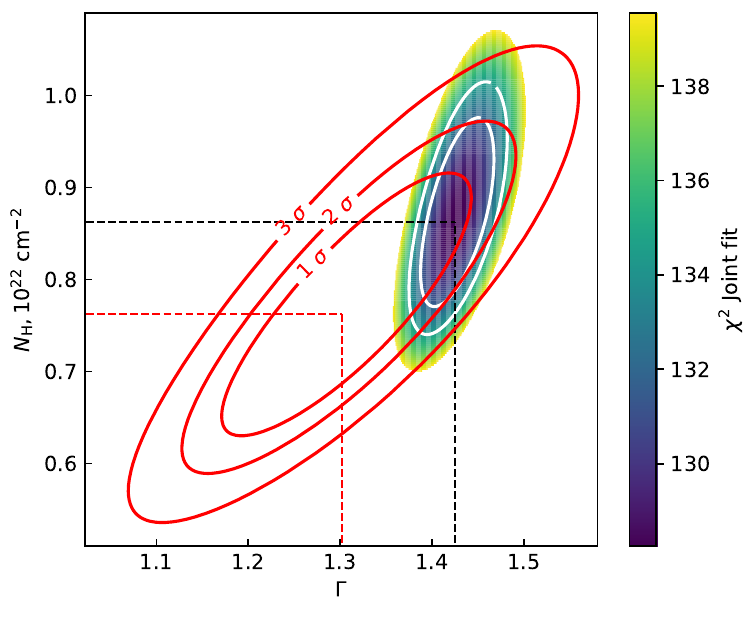}
      \caption{Chi-square ($\chi^2$) maps in the $\Gamma$–$N_{\mathrm{H}}$ parameter space. The color scale represents the joint fit $\chi^2$ values, with regions below $3\sigma$ significance shown. White contours indicate the $1\sigma$ and $2\sigma$ confidence levels. Red contours correspond to $\chi^2$ levels from XRT-only fits (see labels). The analysis includes data from datasets $4-9$ only.
      }
         \label{fig:chi2}
   \end{figure}

In Figure \ref{fig:chi2}, we demonstrate $\Delta \chi^2$ maps in $\Gamma-N_\mathrm{H}$ parameter space from the joint \xrt-\isgri analysis of high-quality datasets 4--9 (corresponding to 17-33 days after periastron) and compare them to the analysis of the corresponding \xrt-only observations. The colorbar represents the change in $\chi^2$ for the given $(N_H, \Gamma)$ relative to the best-fit $\chi^2$ value. The white contours show $1\sigma$ and $2\sigma$ statistical confidence regions, $\Delta\chi^2=3.5$ and $\Delta\chi^2=6.2$ correspondingly, for a joint fit, with the best-fit values of $\Gamma = 1.42 \pm 0.03, N_\mathrm{H}=0.86_{-0.09}^{+0.12}$. The red ellipses show the same confidence regions for the \xrt-only analysis around the best-fit values $\Gamma = 1.30 \pm 0.14, N_\mathrm{H} = 0.76_{-0.13}^{+0.16}$. One can see that the inclusion of \isgri data significantly improves the constraints on the spectral slope. 
 Within error bars, the PL spectral index and hydrogen column density for datasets 4--9 remained consistent with the constant joint fit values (see red shaded regions in Fig.~\ref{fig:LC_noTh}).  

\begin{figure*}[h]
   \centering
   \includegraphics[width=\hsize]{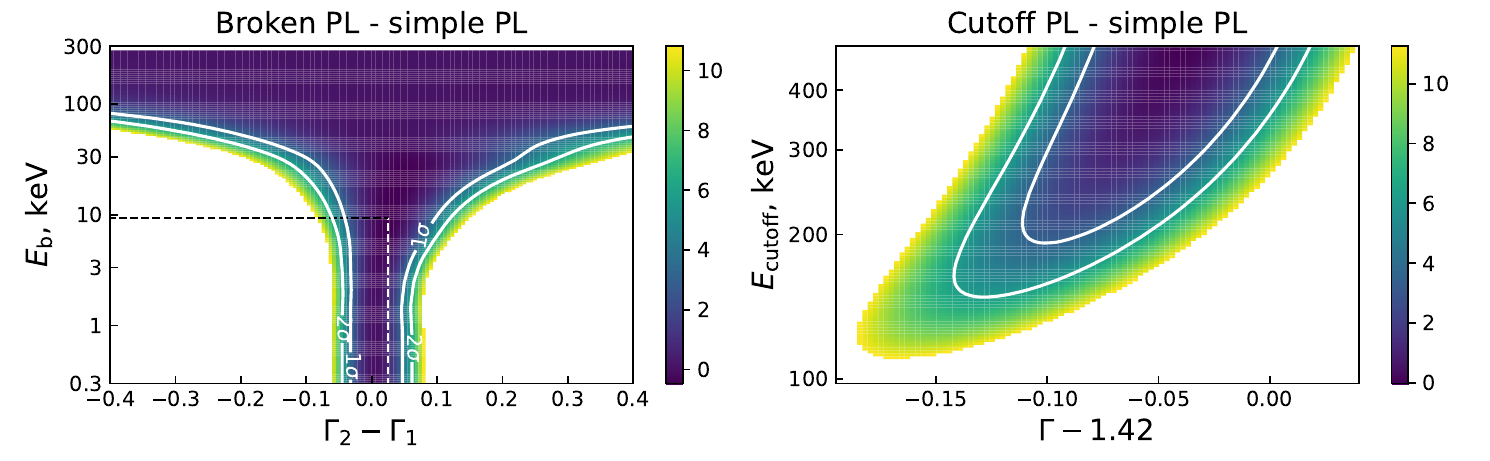}
      \caption{The difference in total $\chi^2$ between a modified PL model and a simple PL model. Left: the modified model is a broken PL, with the map shown in coordinates of break energy $E_{\mathrm{b}}$ and index difference $\Delta \Gamma = \Gamma_2 - \Gamma_1$. The first index was fixed at $\Gamma_1=1.42$, while the second index $\Gamma_2$ was varied. Right: the modified model is an exponential cutoff PL, with the map shown in coordinates of the cutoff energy $E_{\mathrm{cutoff}}$ and spectral index $\Gamma$. The white solid contours indicate significance levels of $1\sigma$ and $2\sigma$. }
         \label{fig:bkpo}
\end{figure*}

Motivated by the considerable refinement of the joint \xrt-\isgri analysis, we performed an additional search for deviations from a simple PL model. We searched for a break in the spectrum by fitting it with a broken PL model:
\begin{align} 
& dN/dE = \left\{ \begin{aligned} 
  E\leq E_\mathrm{b}:\quad N_0 (E/E_\mathrm{b})^{-\Gamma_1},\\
  E>E_\mathrm{b}:\quad N_0 (E/E_\mathrm{b})^{-\Gamma_2}
\end{aligned} \right.
\end{align}
and tested for the spectrum curvature using the exponential cutoff PL:
\begin{align} 
& dN/dE = N_0 (E/1\,{\rm keV})^{-\Gamma} \rm exp(-E/E_{\rm c}). 
\end{align}

The results are shown in Figure ~\ref{fig:bkpo} as the change in $\chi^2$ between the modified and simple PL models. The left panel corresponds to 
the broken PL model,
while the right 
panel to the exponential cutoff PL model.
The contours denote the $1\sigma$ and $2\sigma$ statistical significance contours. 
For the the broken PL model, we additionally fixed $\Gamma_1 = 1.42$ to its best-fit value derived from the joint \xrt--\isgri fit. We explicitly verified that the best improvement of $\delta\chi^2$ for both $\Gamma_1$ and $\Gamma_2$ indexes free is $\textrm{min}(\delta \chi^2) = -2.2$, which remains within the statistical fluctuations of the data.

This figure demonstrates that the data limit spectral breaks to $E_\mathrm{b} \gtrsim 10 $~keV (for $\Delta\Gamma >0.1$) and $E_\mathrm{b} \gtrsim 30$~keV (for $\Delta\Gamma >0.3$) at the $2\sigma$ confidence level. The data do not constrain the break energy above $E_\mathrm{b}\gtrsim 100$~keV. Similarly, we place a lower limit on the cutoff energy of $E_\mathrm{cutoff} \gtrsim 150$~keV at 95\% ($2\sigma$) statistical significance. As before, the $1\sigma$ and $2\sigma$ levels were defined as $\Delta \chi^2 = 3.5$ and $\Delta \chi^2 =6.2$, respectively, above the best-fit $\chi^2$ value. 

\begin{figure}[h]
   \centering
   \includegraphics[width=\hsize]{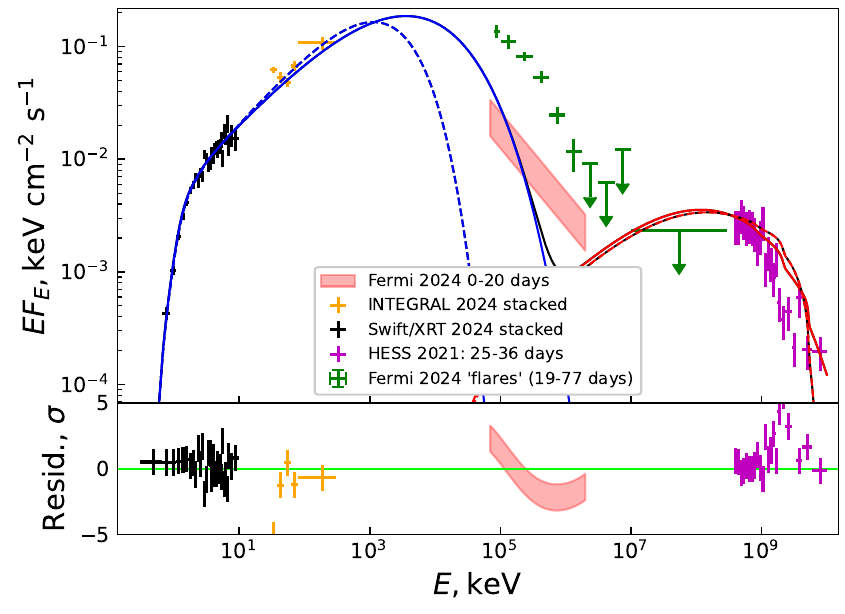}
      \caption{Top: SED of PSR B1259-63. Black: \textit{Swift}/XRT stacked spectrum for the datasets 4--9. Orange: \textit{INTEGRAL}/ISGRI stacked spectrum for the same datasets. Red bound: Fermi 2024 spectrum $0-20$ days after periastron. Green: Fermi spectrum during gamma-ray flares: $19-77$ days after periastron. All Fermi data is from \citet{Che25}. Magenta: 2021 H.E.S.S. spectrum $25-36$ days after periastron from \citet{Hess24}. The lines show the models for synchrotron (blue), inverse Compton (red), and total (black) radiation. Solid (dashed) lines correspond to the $\beta=1\,\,(\beta = 3)$ super-exponential cutoff index. Bottom: residuals for the model spectrum with $\beta=1$.}
         \label{fig:Spec_all}
   \end{figure}

The multiwavelength (keV to TeV) spectral energy distribution (SED) of \psrb  is shown in Fig.~\ref{fig:Spec_all}. The X-ray data correspond to the joint \xrt and \isgri  spectrum from datasets 4--9; the GeV data are the 
Fermi/LAT spectra from~\citet{Che25}, obtained for  different time intervals around the 2024 periastron passage; and the TeV data are \hess results from the 2021 periastron campaign~\citep{Hess24} averaged over 25-36 days after periastron. 

We fit the data with a simple one-zone leptonic model in which the X-ray emission originates from the synchrotron emission of a population of relativistic electrons, while the TeV emission arises from Inverse Compton (IC) emission from the same population.
We assumed that the population of electrons is characterised by a super-exponential cutoff PL spectrum, 
\begin{equation}
    \frac{\mathrm{d}N_\mathrm{e}}{\mathrm{d}\gamma} \propto \gamma^{-p} \exp{\left(-\left(\frac{\gamma}{\gamma_\mathrm{cutoff}}\right)^\beta\right)}.    
\end{equation}

The key parameters of this electron distribution can be estimated analytically based on the X-ray analysis presented above. Namely, the electrons' distribution index is given by $p = 2\Gamma - 1 \approx 1.8$. The derived limits on the break/cutoff energy of the photons' spectrum translate into the constraints for the corresponding quantities in electrons' spectrum via $E_\mathrm{cutoff/b}  \approx 52\,\mbox{keV}\, (B/1\,\mbox{G})\, (E_\mathrm{cutoff/b,~e}/1\,\mbox{TeV})^2$. Thus, we constrain the possible break in the spectrum of electrons to be at $E_\mathrm{b,~e} \gtrsim 0.8$~TeV and cutoff energy to $E_\mathrm{cutoff,~e} \gtrsim 1.7$~TeV. This value is consistent with the cutoff upper limit $E_\mathrm{cutoff} \gtrsim 27$~TeV reported by \hess~\citep{Hess24}.  

We further performed detailed spectral analysis of the discussed model with the \texttt{Naima} package \citep{Zab15}, which uses synchrotron and isotropic IC emission models described in~\citep{Ahar10, Khan14}.

The blue and red lines in Fig.~\ref{fig:Spec_all} show the synchrotron and inverse Compton (IC) components of the spectrum. We explicitly assumed that the GeV data is explained by a separate component, such as a combination of a bremsstrahlung and IC emission from a population of the relativistic electrons of the pulsar wind, \citep[see, e.g.,][]{KhA11, Che20}. Indeed, we see that the synchrotron emission is not enough to explain GeV data taken during Fermi flares. 
Solid and dashed lines correspond to different choices of the super-exponential cutoff PL index $\beta$, namely solid lines correspond to $\beta=1$, while dashed lines correspond to a sharp cutoff $\beta=3$. 

To account for the absorption of the soft X-rays due to interstellar hydrogen, we explicitly convolved the derived photon spectrum with \texttt{Tbabs} absorption model using $N_\mathrm{H}=0.76\times10^{22}\,\textrm{cm}^{-2}$, identical to the value used to fit \xrt data.

The seed photons for the IC scattering were assumed to originate from the optical star, with a temperature of $T_\mathrm{opt} = 3\times 10^4$ K \citep{Neg11} at a distance $r \approx 2.5$ au (at $32$ days after the periastron passage). The $\gamma-\gamma$ absorbtion was not accounted for.

The joint fit of the X-ray and TeV data ranges allowed us to constrain the magnetic field in the emission zone to $B \approx 2$~G. \citet{Che20} preformed a simultaneous modeling of X-ray, GeV, and TeV spectra, obtaining a magnetic field approximately $5-10$ times weaker. However, their analysis was based on 2017 observational data, when the X-ray light curve differed from the 2024 observations.

This value is consistent with the expected toroidal magnetic field for a pulsar \citep{KC84b, TA97}:
\begin{equation}
    B \approx 3\sqrt{\frac{L_\mathrm{spin-down}}{r^2c} \sigma} = 0.9 \, \left(\frac{r_\mathrm{PE}}{{0.1\textrm{~au}}}\right)^{-1}\, \sigma_{0.01}^{1/2} \textrm{~G},
\end{equation}
where $L_\mathrm{spin-down} = 8\times 10^{35} \, \mathrm{erg}~\mathrm{s}^{-1}$ is the pulsar spin-down luminosity, $\sigma \approx 10^{-3}-10^{-1}$ is the magnetization of the pulsar wind, and $r_\mathrm{PE}$ is a distance from the pulsar to the emission zone. The factor of $3$ is due to the compression of the isotropic magnetic field at the shock front. 

This estimation may suggest that the emission zone is located close to the pulsar. The theoretical X-ray light curve modeling by \citet{Chen19} successfully reproduced the observed light curve while employing the equation above, though without the factor of $3$.

We note that a magnetic field of the same order may originate from the optical star. We estimate the stellar magnetic field following \citet{UM92} \citep[see also][]{DeBeck07}:
\begin{equation}
    B_\mathrm{opt} = B_\mathrm{surf}\,\,\frac{V_\infty}{V_\mathrm{rot}}\, \frac{R_\mathrm{opt}^2}{r_\mathrm{A}\,r} \approx 3 \left(\frac{B_\mathrm{surf}}{100\mathrm{~G}}\right) \left(\frac{r_\mathrm{PS}}{2.5\textrm{~au}}\right)^{-1}  \mathrm{G}.
\end{equation}
Here, we adopt the values used by \citet{DeBeck07}: the terminal wind velocity $V_\infty = 3\,V_\mathrm{esc}$, the star rotation velocity $V_\mathrm{rot} = 0.7\, V_\mathrm{esc}$, the Alfvén radius $r_\mathrm{A} = \,3R_\mathrm{opt}$, the star radius $R_\mathrm{opt} = 10\,R_{\odot}$ \citep{MiJ18}, and $r_\mathrm{PS}$ is a binary separation: the distance between the pulsar and the optical star. $B_\mathrm{surf}$ is the magnetic field at the surface of the star. The estimated magnetic field strength is of the same order as the fitted value of $\sim 2~\textrm{G}$. However, no magnetic field has been detected in Be stars, with upper limits on the surface field estimated at $50-100$~G \citep{Wade14}. Moreover, theoretical modeling suggests that a surface field of $100$~G would likely disrupt the decretion disk, with a limit of $B_\mathrm{surf} \lesssim 10$~G  for the strongest surface field that would not completely destroy the disk \citep{udDou18}. 

\section{Conclusions}\label{sec:concl}
We conducted spectral analysis in the hard X-ray range for the 2024 periastron passage of PSR B1259-63 using \textit{INTEGRAL} data. Spectral modeling shows that the data are satisfactorily described by an absorbed PL model in the $0.3-300$ keV energy range. The stacked spectra for $\sim 18-32$ days after periastron yield a photon index of $\Gamma = 1.42 \pm 0.03$.  

At $2\sigma$ significance level, we constrain the spectral break energy to $E_\mathrm{b}\gtrsim 10$ keV (for $\Delta \Gamma > 0.1$) and $E_\mathrm{b}\gtrsim 30$ keV (for $\Delta \Gamma > 0.3$). At the same significance, we exclude an exponential cutoff at energies $E_\mathrm{cutoff}<150$ keV. For the possible break or cutoff in a spectrum of electrons, these results translate to $E_\mathrm{b, e} > 0.4$ TeV (for $\Delta \Gamma > 0.1$) and  $E_\mathrm{b, e} > 0.8$ TeV (for $\Delta \Gamma > 0.3$), and $E_\mathrm{cutoff, e}>1.7$ TeV.

We also compared the theoretical one-zone emission model to the combined observed X-ray and TeV spectrum, incorporating the 2021 H.E.S.S. TeV spectral data. The model with a magnetic field of $\sim 2$~G provides the best fit. Such magnetic field strength at $\sim 30$ days after the periastron passage suggests either that the emission zone lie very close to the pulsar or that the magnetic field could originate from the optical star surface.
\begin{acknowledgements}
      This work was supported by Deutsches Zentrum f\"ur Luft- und Raumfahrt e.V. (DLR) grant 50OR2409.
      BvS acknowledges support from the National Research Foundation of South Africa (grant number 119430). MCh acknowledges support from the European Space Agency (ESA) in the framework of the PRODEX Programme (PEA
4000120711). The authors acknowledge support by the state of Baden-W\"urttemberg through bwHPC.  We acknowledge the use of public data from the Swift data archive. This work is based on observations with INTEGRAL, an ESA project with instruments and science data centre funded by ESA member states.
\end{acknowledgements}

\bibliographystyle{aa} 
\bibliography{Mybib} 

\begin{thebibliography}{26}
\expandafter\ifx\csname natexlab\endcsname\relax\def\natexlab#1{#1}\fi

\bibitem[{{Aharonian} {et~al.}(2010){Aharonian}, {Kelner}, \& {Prosekin}}]{Ahar10}
{Aharonian}, F.~A., {Kelner}, S.~R., \& {Prosekin}, A.~Y. 2010, \prd, 82, 043002

\bibitem[{{Burrows} {et~al.}(2005){Burrows}, {Hill}, {Nousek}, {Kennea}, {Wells}, {Osborne}, {Abbey}, {Beardmore}, {Mukerjee}, {Short}, {Chincarini}, {Campana}, {Citterio}, {Moretti}, {Pagani}, {Tagliaferri}, {Giommi}, {Capalbi}, {Tamburelli}, {Angelini}, {Cusumano}, {Br{\"a}uninger}, {Burkert}, \& {Hartner}}]{XRT05}
{Burrows}, D.~N., {Hill}, J.~E., {Nousek}, J.~A., {et~al.} 2005, \ssr, 120, 165

\bibitem[{{Chen} {et~al.}(2019){Chen}, {Takata}, {Yi}, {Yu}, \& {Cheng}}]{Chen19}
{Chen}, A.~M., {Takata}, J., {Yi}, S.~X., {Yu}, Y.~W., \& {Cheng}, K.~S. 2019, \aap, 627, A87

\bibitem[{{Chernyakova} \& {Malyshev}(2020)}]{ChM20}
{Chernyakova}, M. \& {Malyshev}, D. 2020, in Multifrequency Behaviour of High Energy Cosmic Sources - XIII. 3-8 June 2019. Palermo, 45

\bibitem[{{Chernyakova} {et~al.}(2020){Chernyakova}, {Malyshev}, {Mc Keague}, {van Soelen}, {Marais}, {Martin-Carrillo}, \& {Murphy}}]{Che20}
{Chernyakova}, M., {Malyshev}, D., {Mc Keague}, S., {et~al.} 2020, \mnras, 497, 648

\bibitem[{{Chernyakova} {et~al.}(2025){Chernyakova}, {Malyshev}, {van Soelen}, {Finn Gallagher}, {Matchett}, {Russell}, {van den Eijnden}, {Lower}, {Johnston}, {Tsygankov}, {Salganik}, \& {Shebalkova}}]{Che25}
{Chernyakova}, M., {Malyshev}, D., {van Soelen}, B., {et~al.} 2025, \mnras, 536, 247

\bibitem[{{Chernyakova} {et~al.}(2024){Chernyakova}, {Malyshev}, {van Soelen}, {Mc Keague}, {O'Sullivan}, \& {Buckley}}]{Che24}
{Chernyakova}, M., {Malyshev}, D., {van Soelen}, B., {et~al.} 2024, \mnras, 528, 5231

\bibitem[{{Chernyakova} {et~al.}(2021){Chernyakova}, {Malyshev}, {van Soelen}, {O'Sullivan}, {Sobey}, {Tsygankov}, {Mc Keague}, {Green}, {Kirwan}, {Santangelo}, {P{\"u}hlhofer}, \& {Monageng}}]{Che21}
{Chernyakova}, M., {Malyshev}, D., {van Soelen}, B., {et~al.} 2021, Universe, 7, 242

\bibitem[{{Chernyakova} {et~al.}(2015){Chernyakova}, {Neronov}, {van Soelen}, {Callanan}, {O'Shaughnessy}, {Babyk}, {Tsygankov}, {Vovk}, {Krivonos}, {Tomsick}, {Malyshev}, {Li}, {Wood}, {Torres}, {Zhang}, {Kretschmar}, {McSwain}, {Buckley}, \& {Koen}}]{Che15}
{Chernyakova}, M., {Neronov}, A., {van Soelen}, B., {et~al.} 2015, \mnras, 454, 1358

\bibitem[{{De Becker}(2007)}]{DeBeck07}
{De Becker}, M. 2007, \aapr, 14, 171

\bibitem[{{H.~E.~S.~S. Collaboration} {et~al.}(2024){H.~E.~S.~S. Collaboration}, {Aharonian}, {Ait Benkhali}, {Aschersleben}, {Ashkar}, {Backes}, {Barbosa Martins}, {Batzofin}, {Becherini}, {Berge}, {Bernl{\"o}hr}, {B{\"o}ttcher}, {Boisson}, {Bolmont}, {de Bony de Lavergne}, {Borowska}, {Bouyahiaoui}, {Brose}, {Brown}, {Brun}, {Bruno}, {Bulik}, {Burger-Scheidlin}, {Caroff}, {Casanova}, {Celic}, {Cerruti}, {Chand}, {Chandra}, {Chen}, {Chibueze}, {Chibueze}, {Cotter}, {Damascene Mbarubucyeye}, {Devin}, {Djuvsland}, {Dmytriiev}, {Egberts}, {Einecke}, {Ernenwein}, {Fontaine}, {Funk}, {Gabici}, {Gallant}, {Glawion}, {Glicenstein}, {Goswami}, {Grolleron}, {Haerer}, {He{\ss}}, {Hofmann}, {Holch}, {Holler}, {Huang}, {Jamrozy}, {Jankowsky}, {Joshi}, {Jung-Richardt}, {Kasai}, {Katarzy{\'n}ski}, {Khangulyan}, {Khatoon}, {Kh{\'e}lifi}, {Klu{\'z}niak}, {Komin}, {Kosack}, {Kostunin}, {Kundu}, {Lang}, {Le Stum}, {Leitl}, {Lemi{\`e}re}, {Lemoine-Goumard}, {Lenain}, {Leuschner}, {Mackey}, {Malyshev}, {Mart{\'\i}-Devesa},
  {Marx}, {Mehta}, {Meintjes}, {Mitchell}, {Moderski}, {Mohrmann}, {Montanari}, {Moulin}, {Murach}, {de Naurois}, {Niemiec}, {Ohm}, {de Ona Wilhelmi}, {Ostrowski}, {Panny}, {Panter}, {Parsons}, {Pensec}, {Peron}, {Prokhorov}, {P{\"u}hlhofer}, {Punch}, {Quirrenbach}, {Regeard}, {Reimer}, {Reimer}, {Reis}, {Ren}, {Rieger}, {Rudak}, {Ruiz-Velasco}, {Sahakian}, {Salzmann}, {Santangelo}, {Sasaki}, {Sch{\"a}fer}, {Sch{\"u}ssler}, {Schutte}, {Shapopi}, {Spencer}, {Stawarz}, {Steenkamp}, {Steinmassl}, {Steppa}, {Streil}, {Sushch}, {Takahashi}, {Tanaka}, {Taylor}, {Terrier}, {Thorpe-Morgan}, {Tluczykont}, {Unbehaun}, {van Eldik}, {van Soelen}, {Vecchi}, {Venter}, {Vink}, {Wach}, {Wagner}, {Werner}, {Wierzcholska}, {Zacharias}, {Zdziarski}, {Zech}, \& {{\.Z}ywucka}}]{Hess24}
{H.~E.~S.~S. Collaboration}, {Aharonian}, F., {Ait Benkhali}, F., {et~al.} 2024, \aap, 687, A219

\bibitem[{{Johnston} {et~al.}(1992){Johnston}, {Manchester}, {Lyne}, {Bailes}, {Kaspi}, {Qiao}, \& {D'Amico}}]{John92}
{Johnston}, S., {Manchester}, R.~N., {Lyne}, A.~G., {et~al.} 1992, \apjl, 387, L37

\bibitem[{{Kennel} \& {Coroniti}(1984)}]{KC84b}
{Kennel}, C.~F. \& {Coroniti}, F.~V. 1984, \apj, 283, 694

\bibitem[{{Khangulyan} {et~al.}(2011){Khangulyan}, {Aharonian}, {Bogovalov}, \& {Rib{\'o}}}]{KhA11}
{Khangulyan}, D., {Aharonian}, F.~A., {Bogovalov}, S.~V., \& {Rib{\'o}}, M. 2011, \apj, 742, 98

\bibitem[{{Khangulyan} {et~al.}(2014){Khangulyan}, {Aharonian}, \& {Kelner}}]{Khan14}
{Khangulyan}, D., {Aharonian}, F.~A., \& {Kelner}, S.~R. 2014, \apj, 783, 100

\bibitem[{{Melatos} {et~al.}(1995){Melatos}, {Johnston}, \& {Melrose}}]{MJM95}
{Melatos}, A., {Johnston}, S., \& {Melrose}, D.~B. 1995, \mnras, 275, 381

\bibitem[{{Miller-Jones} {et~al.}(2018){Miller-Jones}, {Deller}, {Shannon}, {Dodson}, {Mold{\'o}n}, {Rib{\'o}}, {Dubus}, {Johnston}, {Paredes}, {Ransom}, \& {Tomsick}}]{MiJ18}
{Miller-Jones}, J.~C.~A., {Deller}, A.~T., {Shannon}, R.~M., {et~al.} 2018, \mnras, 479, 4849

\bibitem[{{Negueruela} {et~al.}(2011){Negueruela}, {Rib{\'o}}, {Herrero}, {Lorenzo}, {Khangulyan}, \& {Aharonian}}]{Neg11}
{Negueruela}, I., {Rib{\'o}}, M., {Herrero}, A., {et~al.} 2011, \apjl, 732, L11

\bibitem[{{Salganik} {et~al.}(2025){Salganik}, {Tsygankov}, {Chernyakova}, {Malyshev}, \& {Poutanen}}]{Sal25}
{Salganik}, A., {Tsygankov}, S.~S., {Chernyakova}, M., {Malyshev}, D., \& {Poutanen}, J. 2025, arXiv e-prints, arXiv:2504.11263

\bibitem[{{Shaw} {et~al.}(2004){Shaw}, {Chernyakova}, {Rodriguez}, {Walter}, {Kretschmar}, \& {Mereghetti}}]{Shaw04}
{Shaw}, S.~E., {Chernyakova}, M., {Rodriguez}, J., {et~al.} 2004, \aap, 426, L33

\bibitem[{{Tavani} \& {Arons}(1997)}]{TA97}
{Tavani}, M. \& {Arons}, J. 1997, \apj, 477, 439

\bibitem[{{Tavani} {et~al.}(1994){Tavani}, {Arons}, \& {Kaspi}}]{TAK94}
{Tavani}, M., {Arons}, J., \& {Kaspi}, V.~M. 1994, \apjl, 433, L37

\bibitem[{{ud-Doula} {et~al.}(2018){ud-Doula}, {Owocki}, \& {Kee}}]{udDou18}
{ud-Doula}, A., {Owocki}, S.~P., \& {Kee}, N.~D. 2018, \mnras, 478, 3049

\bibitem[{{Usov} \& {Melrose}(1992)}]{UM92}
{Usov}, V.~V. \& {Melrose}, D.~B. 1992, \apj, 395, 575

\bibitem[{{Wade} {et~al.}(2014){Wade}, {Petit}, {Grunhut}, \& {Neiner}}]{Wade14}
{Wade}, G.~A., {Petit}, V., {Grunhut}, J., \& {Neiner}, C. 2014, arXiv e-prints, arXiv:1411.6165

\bibitem[{{Zabalza}(2015)}]{Zab15}
{Zabalza}, V. 2015, in International Cosmic Ray Conference, Vol.~34, 34th International Cosmic Ray Conference (ICRC2015), 922

\end{thebibliography}

\end{document}